\documentclass[11pt]{article}
\usepackage{float}
\usepackage[caption = false]{subfig}

\usepackage{hyperref}
\usepackage{longtable}
\usepackage{bookmark}
\usepackage{booktabs}

\usepackage{amssymb}
\usepackage{color}
\usepackage{amsmath}
\usepackage{amsfonts}
\usepackage{graphicx}
\usepackage{multirow}

\usepackage[a4paper,margin=1in]{geometry}
\usepackage{caption}
\title{Analytic Pricing of Bermudan Swaptions with Few Exercise Dates }

\author{Emiliano Papa\thanks{Quant Analyst, London. Email:  e.papa1@yahoo.com} 
\\
}

\date{\today}

\newcommand{\be}{\begin{equation}}
\newcommand{\ee}{\end{equation}}

\newcommand{\bea}{\begin{eqnarray}}
\newcommand{\eea}{\end{eqnarray}}

\newcommand{\rd}{{\rm d}}

\begin{document}

\maketitle

 \abstract{
In this paper, we consider pricing a Bermudan swaption with a small number of exercise dates.
We begin with the case of two exercise dates. 
  
In this limit, 
we show that the Bermudan price decomposes into the sum of short-dated European swaptions, setting an upper bound, minus a correction term. This correction is expressed as an integral involving a forward volatility agreement type payoff with start at the first exercise date, and it can be evaluated in closed form. 
The magnitude of the correction is smaller when variance is front loaded and larger when it is back-loaded.

We extend to three-exercise Bermudans via backward induction under rolling forward measures. A key feature is boundary linearity
enabling further analytic steps.

The exercise boundary of options splits into a strike-dependent term and a variance term; 
together they determine optimal exercise. The linear term is negative, supressing the exponentials in subsequent steps and aiding analytic calculations.

This boundary linearity extends to multiple exercise dates and yields pricing formulas with the same decomposition, showing how optionality accumulates across exercise dates.

We conclude that the Bermudan can be reconstructed by adding, at each exercise date, the initial short swaption with an increasingly  higher strike and subtracting the integrated payoffs of all forward-starting receiver swaptions starting at that date. The corresponding double and higher-order integrals decrease rapidly and, in the presence of only a few exercise dates, can be safely neglected without materially impacting the valuation. The general case is discussed at the end. 
}

\section{Two-Exercise Bermudan Swaption Pricing}

We work in the short-rate framework, assuming an instantaneous risk free rate $r(t)$ and the corresponding bank account numeraire $B(t)=\exp\{\int_0^t r(u) \rd u\}$. With this numeraire it is associated the Q-risk-neutral measure, which is equivalent to a  forward measure, convenient choice for carrying out calculations in this paper. 
We work in the Hull-White model, for which the dynamics of the bonds in the $t$-forward measure\cite{AndersenPiterbarg2010}-\cite{AndersenPiterbarg2010_III} is
\be
B(t,T) = \frac{B(0,T)}{B(0,t)} \exp\left\{- \int_0^t \Bigl(\sigma(u,T) -\sigma(u,t) \Bigr)\rd W^t(u) -\frac{1}{2} \int_0^t \Bigl(\sigma(u,T) -\sigma(u,t) \Bigr)^2 \rd u\right\}
\;.
\label{Bond_Reconstruction}
\ee
The first problem we consider is a Bermudan swaption with two call dates (see %Henrard 
\cite{Henrard}, \cite{Feldman} and references therein),
\bea
V_0 = 
B(0,\theta_1)
E^{\theta_1} \left[ \left. {\rm max} \Bigl( {\rm Swap}_2(\theta_1),  {\rm {\rm Swaption}_1}(\theta_1)  \Bigr)  \right|{\cal F}_0\right]
\quad.
\label{Berm01}
\eea
We consider a time horizon spanned by points, $T_i$ equally spaced between $T_0$, $\cdots$, $T_n$.
At $\theta_1$ we compare the prices of the Swap starting at $T_{1,0}$ and ending at $T_{1,n_1}$ versus the price of the Swaption on the underlying Swap that starts at $T_{2,0}$ and ends at $T_{2,n_2}$, where $\theta_i \le T_{i0} < T_{i1}\ldots < T_{i n_i}$, for $i=1,2$.

At $\theta_1$ the price of the $\theta_2$-expiry Swaption is given by
\bea
{\rm Swaption}_1(\theta_1) = 
B(\theta_1,\theta_2) E^{\theta_2}\left[ \left. \left( \sum_{i=0}^{n_2}  c_{2,i} B(\theta_2,T_{2,i}; x(\theta_1, \theta_2;\theta_2)) \right)^+ \right|{\cal F}_{\theta_1}\right]
\quad,
\label{Swation2_theta2}
\eea
where we have 
$c_{2,0}=1$, 
$c_{2,i} = - \tau_{2,i} K$, 
$c_{2,n_2} = -(1+ \tau_{2,n_2} K)$.

We introduce the Gaussian $x\in N(0,1)$ in the $\theta_2$-forward measure in the interval $[\theta_1,\theta_2]$. 
The corresponding variance of the process is
$\alpha^2_{i}(\theta_1, t;\theta_2) = \int_{\theta_1}^t  \sigma^2(u)\Bigl( G(u,T_{i}) - G(u,\theta_2) \Bigr)^2 \rd u$,
which for $\alpha_i(\theta_1,\theta_2;\theta_2)$ we write as $\alpha_i(\theta_1,\theta_2)$.

For the Bermudan price we discount  from $\theta_1$ under the $\theta_1$-forward measure
{\small
\bea
V_0 = 
B(0,\theta_1)
E^{\theta_1} && 
\hspace{-6mm}
\left[ {\rm max} 
\left( 
\sum_{i=0}^{n_1}  c_{1,i} 
\frac{ B(0,T_{1,i}) }{ B(0,\theta_1)} 
\exp\left\{ - \alpha_{1,i}(\theta_1) y - \frac{1}{2} \alpha^2_{1,i}(\theta_1)    \right\} 
,
\right.\right.
\\[3mm]\nonumber
&&
\hspace{-20mm}
\left.\left.
\sum_{i=0}^{n_2}  c_{2,i} 
\frac{ B(0,T_{2,i}) }{ B(0,\theta_1) } 
\exp\left\{ - \alpha_{2,i}(\theta_1) y - \frac{1}{2} \alpha^2_{2,i}(\theta_1)    \right\} 
N\Bigl(-x^*(y) - \alpha_{2,i}(\theta_1, \theta_2)\Bigr) 
\right)   \right]
\;.
\eea
}
In this equation $y$ is a standard normal distribution $N(0,1)$ for the evolution of the Bond in the interval $[0,\theta_1]$ in the $\theta_1$-forward measure.

Here, $x^*$ can be written as a function of $B(\theta_1,T_{2,i})$, but these quantities are not known explicitly at $\theta_1$, instead depend on $y$. Consequently, $x^*$ varies with $y$, and its dependence can be obtained numerically by solving,\footnote{
\be
\label{x*}
 \sum_{i=0}^{n_2} c_{2,i} B(0, T_{2,i}) \exp\left\{ 
- \alpha_{2, i}(\theta_1) y - \frac{1}{2} \alpha^2_{2, i}(\theta_1)
-\alpha_{2,i}(\theta_1, \theta_2) x^* - \frac{1}{2} \alpha_{2, i}^2(\theta_1, \theta_2) \right\}
= 0
\quad.
\ee
To first order we get an approximate analytic solution
\be
x^*(y) \approx \frac{  
   \sum_{i=0}^{n_2}  c_{2,i} B(0,T_{2,i}) 
   \exp\Bigl\{ - \alpha_{2,i}(\theta_1) y - \frac{1}{2} \alpha^2_{2,i}(\theta_1)   - \frac{1}{2} \alpha^2_{2,i}(\theta_1, \theta_2)  \Bigr\}  
   }{  
   \sum_{i=0}^{n_2}  c_{2,i} \alpha_{2,i}(\theta_1, \theta_2)   B(0,T_{2,i}) 
   \exp\Bigl\{ - \alpha_{2,i}(\theta_1) y - \frac{1}{2} \alpha^2_{2,i}(\theta_1) - \frac{1}{2} \alpha^2_{2,i}(\theta_1, \theta_2)\Bigr\}
}
\quad.
\ee
In the zero strike case, it simplifies (note the negative sign of the linear factor),
\bea
x^*(y) &=& \frac{1}{ (\alpha_{2,n_2} - \alpha_{2,0})(\theta_1, \theta_2)}
\left[
\ln B(0;T_{2,0},T_{2,n_2}) -\frac{1}{2} \left(\alpha^2_{2,n_2} - \alpha^2_{2,0}\right)(\theta_1, \theta_2)
-\frac{1}{2} \left(\alpha^2_{2,n_2} - \alpha^2_{2,0}\right)(\theta_1)
\right]
\nonumber\\[3mm]
&-&
\frac{\left(\alpha_{2,n_2} - \alpha_{2,0}\right)(\theta_1)}{(\alpha_{2,n_2} - \alpha_{2,0})(\theta_1, \theta_2)} \; y \quad.
\eea
} 
for $x^*(y)$.
Further analytical progress is hindered by the complicated structure of 
$x^*(y)$. 
The problem becomes tractable in two cases: for longer maturities with zero strike, or for a one-period swaption. In both settings, the expression reduces to a linear function of the state variable $y$. 
The latter corresponds to the final swaption encountered in the backward-start valuation of the Bermudan. As the recursion proceeds backward in time, additional exponential terms emerge; however, their contribution is suppressed by the associated damping factors

We begin by considering a Swaption with two exercise dates and the minimum number of underlying swap payment dates, namely three dates $T_0$, $T_1$, and $T_2$. For concreteness, we set $T_0=5$y, $T_1=6$y, and $T_2=7$y.

Using the identity ${\rm max}(a,b) = b + {\rm max}(a-b,0)$, the Bermudan Swaption can be expressed as the sum of ${\rm Swaption}_1$ (the shorter-tenor instrument over $[T_1,T_2])$ and the spread between ${\rm Swap}_2(T_0)$ (over $[T_0,T_2]$) and ${\rm Swaption}_1(T_0)$,
\bea
\label{Berm4}
V_0 = {\rm Swaption}_1(0) + 
B(0,T_0)
E^{T_0} \left[
\left.
{\rm max} \left\{ {\rm Swap}_2(T_0) - {\rm Swaption}_1(T_0), 0 \right\}
\right|{\cal F}_0
\right]
\quad.
\eea
The short $\text{Swaption}_{[T_1,T_2]}(T_0)$
corresponds to the continuation value of the Bermudan at $T_0$. 
The integration is over $x \in (x^*,\infty)$, where the swap on $[T_1,T_2]$ is positive.
Determining $x^*$ requires the value of the forward bond $B(T_0;T_1,T_2)$ at time $T_0$. 
 At this point 
 the exponential term exactly offsets the ratio $B(T_0;T_1,T_2)/\tilde{K}_2$, bringing it to unity; we abbreviate $\tilde{K}_2 =1/(1+\tau_2 K)$.

We switch to the $T_0$-forward measure, under which discounting from $T_0$ to $t=0$ is performed. 
%In this measure, the forward bond $B(T_0;T_1,T_2)$ is given on the right-hand side below, and $x^*(y)$ on the left-hand side
%-Corrected after submission - 25th May 2026:
In this measure, $x^*(y)$ and the forward bond $B(T_0;T_1,T_2)$ are given as follows 
\bea
\label{FwdBond_T0-fwd_000}
x^*_{B(T_0;T_0,T_1)} &=& \frac{1}{\alpha_{2}(T_0, T_1)} \left[ \ln \left( \frac{ B(T_0;T_1,T_2)}{\tilde{K}_2} \right) 
-
\frac{1}{2} \alpha^2_{2}(T_0, T_1)    \right] \quad,
\\[3mm]
B(T_0;T_1,T_2) &=& B(0;T_1,T_2) \frac{\exp\Bigl\{ - \alpha_2(T_0) y - \frac{1}{2} \alpha^2_{2}(T_0)    \Bigr\}   }{ \exp\Bigl\{ - \alpha_1(T_0) y - \frac{1}{2} \alpha^2_{1}(T_0)    \Bigr\}   } 
\quad.
\label{FwdBond_T0-fwd_00}
\eea
The evolution on $[0,T_0]$ is not a martingale for the forward bond, unlike its evolution on $[T_0,T_1]$ under the $T_1$-forward measure.

From Equations~(\ref{FwdBond_T0-fwd_000})--(\ref{FwdBond_T0-fwd_00}),  
we obtain an equation for the exercise boundary $x^*(y)$
in the form $x^*(y) = d_- + by$, and $x^*(y) + \alpha_2(T_0,T_1) = d_+ + by$.
The first component is strike-dependent,  $d_\pm \sim \frac{1}{\alpha_{2}(T_0, T_1)} 
\left[ 
r(\bar{T}_1) - K  \right] (T_2-T_1)$, and measures the moneyness of the last one-period Swaption with Ito-terms offsets. 
The factor $b= - \bigl( \alpha_{2}(T_0) - \alpha_{1}(T_0) \bigr)/ \alpha_{2}(T_0, T_1)$, has no strike dependence, and measures roughly the distribution of the variance of the terminal Swaption  in the interval $[0,T_0]$ versus the interval $[T_0,T_1]$, as $b^2 {\sim} T_0/( T_1 - T_0)$.

The price of the two call date Bermudan Swaption Eq.~(\ref{Berm4}) can be simplified using put-call parity
\be
{\rm Swap}_{[T_0,T_2]}(T_0) - {\rm Swaption}_{[T_1,T_2]}(T_0) = {\rm Swap_{[T_0,T_1]}}(T_0) - {\rm \tilde{S}waption}_{[T_1,T_2]}(T_0)
\quad,
\ee
where ${\rm \tilde{S}waption}_{[T_1,T_2]}(T_0)$ is a receiver swaption. 

We denote the discounted expectation, corresponding to the second term of Eq.~(\ref{Berm4}), by  ${\rm SpreadOpt}_{(2-1)}(0)$. Options within this spread-option reflect presence of forward volatility products.
Equation~(\ref{Berm4}) can then be written explicitly, 
\bea
%\label{Berm06}
V_0 &=& {\rm Swaption}_1(0) + 
B(0,T_0)
E^{T_0} 
\left[
 \left\{ 
 1- 
\frac{ B(0;T_0,T_1) }{ \tilde{K}_1 } 
 \exp
 \Bigl\{ 
 - \alpha_1(T_0) y - \frac{1}{2} \alpha^2_{1}(T_0)    
 \Bigr\}
\right.\right.
\nonumber\\[3mm]\nonumber
&&
%\left.\left.\left.
%\hspace{-10mm}
-
\left( 
\frac{ B(0;T_0,T_2) }{ \tilde{K}_2 } 
\exp\left\{  - \alpha_2(T_0) y - \frac{1}{2} \alpha^2_{2}(T_0)     \right\} 
\right)
N\Bigl(d_+ + b  y\Bigr) 
\\[3mm]
&&
\left.\left.\left.
+
\left(
  \frac{ B(0,T_1) }{ B(0,T_0) } 
 \exp
 \left\{ - \alpha_1(T_0) y - \frac{1}{2} \alpha^2_{1}(T_0) \right\}
 \right)
  N(d_-  + b y)
\right\} ^+ \right|{\cal F}_0
\right] \quad.
\label{Berm06}
\eea
Notice that the receiver swaption is a monotincally decreasing function of $y$, as at higher rates we pay more and receive less. Therefore the subtraction of the receiver swaption maintains the monotonicity of the first swap, it simply moves the root to the right.
The two additional exponential terms on the receiver swaption in the second line contain damping factors of the form $N(d_\pm + by)$, which cause them to decay more rapidly for large positive values of 
$y$, due to the negative sign of the coefficient $b$.
These two factors lead to small contributions of the forward volatility agreement, but also to the small change of the boundary perturbation, from $y_0^*$ to $y^*$.

The expectation leads to a Bermudan price of 
\bea
\label{Berm5_0}
V(0) &=& {\rm Swaption}_{[T_1,T_2]}(0) + {\rm Swaption}_{[T_0,T_1]}(y^*) 
\nonumber\\[3mm]
&-& B(0,T_0)\int_{y^*}^{ \infty} {\rm \tilde{S}waption}\Bigl(T_0;T_1,T_2, x^*(y)\Bigr) \phi(y) \rd y
\quad.
\eea
The integral is taken over a forward starting Swaption. This can be calculated analytically in closed form.

\subsection{Analytic Price for Two-Exercise Bermudans}

The price of the two-exercise Bermudan,  Eq.~(\ref{Berm5_0}), is expressed as the sum of two short swaptions - which serves as an upper bound for the Bermudan - minus a final term consisting of an integral over forward starting ${\rm \tilde{S}waption}\Bigl(T_0;T_1,T_2, y\Bigr)$. 
The term ${\rm Swaption}_{[T_0,T_1]}(y^*)$ is not a swaption with the Bermudan strike, 
(as given by $y_0^*$), since the boundary is shifted to the right, $y^* > y^*_0$. 
Consequently, the resulting upper bound is tighter than the sum of the two market swaptions set at the Bermudan strike level.

 The correction integral in Eq.~(\ref{Berm5_0}) reduces to two $TT$-functions\footnote{
 In Eq.~(\ref{Berm5_0}) 
two integrals of the following form arise,
\be
\int_{y^*}^\infty e^{-\frac{1}{2}\bigl(y+\alpha_\pm(T_0)\bigr)^2}   N( d_\pm  + b  y) \Bigr) \frac{\rd y}{\sqrt{2\pi} }
= \Bigl. TT(y, a_\pm,b)\Bigr|_{y^* +\alpha_\pm}^\infty
\quad,
\label{TT_integrals_0}
\ee
where  
$\alpha_+ = \alpha_2(T_0)$, $\alpha_- = \alpha_1(T_0)$ and denote $y^*_\pm = y^*+\alpha_\pm$ and $a_\pm = (d_\pm - \alpha_\pm b)$; $b$ is negative, and when $b=-1$, the adjustment of the new zero $y^*$ and the strike dependent part of the previous zero is the same $y^*_\pm =y^*+\alpha_\pm$, $\tilde{x}^*_\pm = x^*(0)+\alpha_\pm$, where 
$\tilde{x}^*_\pm=a_\pm$, with 
\be
a_\pm =
\frac{1}{\alpha_{2}(T_0, T_1) }
\left[  
 \ln \left(  \frac{B(0;T_1,T_2)}{\tilde{K}_2} \right)
\pm
\frac{1}{2}
\alpha^2_{2}(T_0, T_1) 
\pm
\frac{1}{2}
\Bigl(\alpha_{2}(T_0) - \alpha_{1}(T_0)   \Bigr)^2
\right]
\quad,
\ee
and the $TT$-functions are given by
\bea
TT(x, a_\pm,b) &=&  
T\left( x, \frac{a_\pm}{x\sqrt{1+b^2}}  \right)
+
T\left( \frac{a_\pm}{\sqrt{1+b^2}}, \frac{x\sqrt{1+b^2}}{a_\pm}  \right)
-
T\left( x, \frac{a_\pm+bx}{x}  \right)
\\[3mm] \nonumber
&-&
T\left( \frac{a_\pm}{\sqrt{1+b^2}}, \frac{a_\pm b+x(1+b^2)}{a_\pm}  \right)
+
N(x) N\left( \frac{a_\pm}{\sqrt{1+b^2}} \right)
\quad.
\label{Analytic_Switch}
\eea
In this equation we use the Owen-T function \cite{OwenT}, with two interesting limits  
\be
T(h,a) =
\frac{1}{2\pi} 
 \int_0^a \frac{ e^{ -\frac{ h^2}{2} (1+x^2) }}{ 1+x^2} \rd x
 \quad, \quad
T(h,\infty) =
\frac{1}{2\pi} 
 \int_0^\infty \frac{ e^{ -\frac{ h^2}{2} (1+x^2) }}{ 1+x^2} \rd x
 = 
 \frac{1}{2 } N(-h)
 \quad. 
 \label{Large_h_Large_a}
\ee 
It vanishes as $h\to \infty$, $T(h\gg 0 ,\infty)  \to 0$, and attains its maximum value 1/4 as $h\to 0$, 
$T(h\to 0 ,\infty) \to \frac{1}{4}$. 
};
collecting all terms for the Bermudan price we get the decomposition 
\bea
\label{two_exercise_Berm}
V(0)
&=&
 B(0,T_1) \left[ N\bigl( -x^* \bigr) -  \frac{B(0;T_1,T_2)}{\tilde{K}_2} N\bigl( -x^*-\alpha_{2}(T_1) \bigr) \right]
\\[3mm] \nonumber
&+&  B(0,T_0)
\left[
N(-y^*)
-\frac{ B(0;T_0,T_1) }{ \tilde{K}_1}  N\bigl( - y^* - \alpha_1(T_0)\bigr) 
\right]
\nonumber
\\[3mm]
\nonumber
&-&
B(0,T_1)\left\{
\frac{ B(0;T_1,T_2) }{ \tilde{K}_2}
\left[ N\left(\frac{\tilde{x}^*_+}{\sqrt{1+b^2}} \right) 
-
 TT\bigl(y^*_+, \tilde{x}^*_+ \bigr)
\right]
\right.
\\[3mm] \nonumber
&&
\left.
-
\left[
N\left(\frac{\tilde{x}^*_-}{\sqrt{1+b^2}} \right) 
-
TT\bigl(y^*_- , \tilde{x}^*_- \bigr)
\right]
\right\}
\quad.
%\label{two_exercise_Berm}
\eea
This expression gives the exact analytic price of the Bermudan, without approximation. The valuation depends on both zeros - the current zero and the previous one.
%($y^*$) and the previous one ($x^*$), slightly modified in the final term to $y^*_\pm$ and $x^*_\pm$.

Note that the factor $(1+b^2)$ scales the strike-dependent terms $\tilde{x}^*_\pm$ (for which we also use the notation $a_\pm$ ), consistent with the intuition that higher front-end variance corresponds to lower effective strikes, and vice versa. Both 
$\frac{ \tilde{x}^*_\pm }{\sqrt{1+b^2}}$ and $y^*$ influence the magnitude of the correction term, with the effect of $y^*$ being dominant. 
\begin{figure}[H] 
\center
{
\hspace{1mm}
\includegraphics[width =0.44\textwidth]{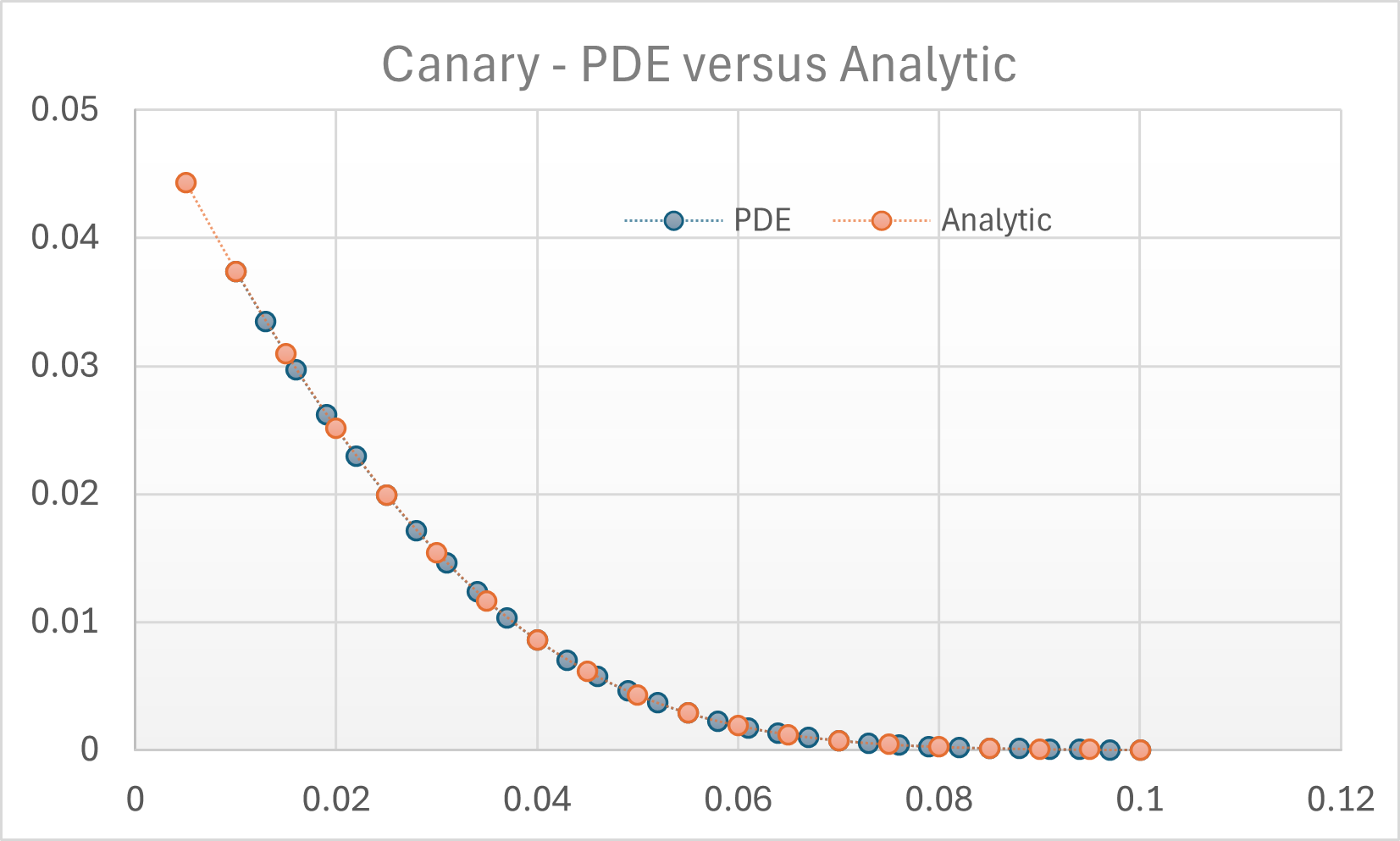}
\includegraphics[width =0.44\textwidth]{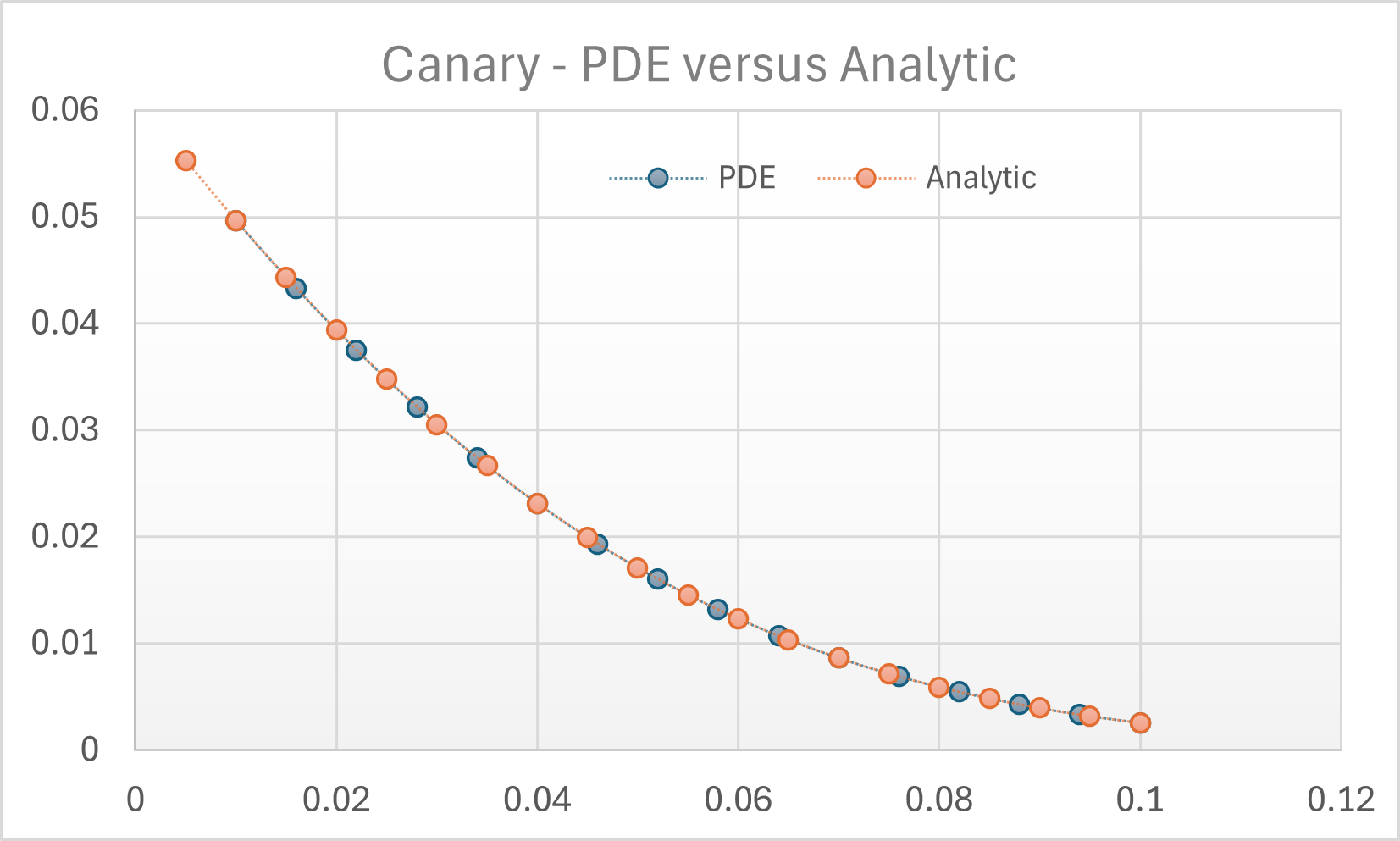}
}
\caption{ PV of two-point Bermudan on an annual Swap starting at $T_0=5$y, ending at $T_2=7$y, with two exercise dates on $T_0=5$y and $T_1=6$y for a range of strikes, are shown. The volatility is taken to be $\sigma=1\%$ on the left and $\sigma=2\%$ on the right.
}
\end{figure}

\hspace{-6mm}
\begin{minipage}{0.5\textwidth} %\hspace{-10mm}
The TT functions $TT(x,a_+, b)$ and $TT(x, a_-, b)$ of Eq.~(\ref{Analytic_Switch}) are very similar to the Gaussian cumulative distribution functions with the difference that they extend from 0 to $N\left( \frac{a_+ }{\sqrt{1+b^2}}\right)$ and $N\left( \frac{a_- }{\sqrt{1+b^2}}\right)$, respectively. 
At the upper integration bound,   
$TT\left( \infty,\frac{a_\pm}{\sqrt{1+b^2}} \right) = N\left( \frac{a_\pm}{\sqrt{1+b^2}} \right)$. 
\end{minipage}
\hfill
\begin{minipage}{0.34\textwidth}
\centering
\hspace{-0mm}
\includegraphics[width=\textwidth]{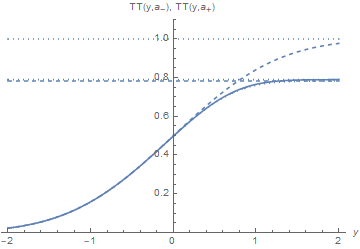}
\captionof{figure}{The functions $TT(x,a_\pm, b)$.} 
\end{minipage}

\begin{table}[hb] 
\caption{
Bermudan, swaption and spread option components are shown, numerical compared with analytical Eq.~(\ref{two_exercise_Berm}). The calculation is using numeric solution for $y^*$. The parameters are  $K=5\%$, $r=3\%$, $\sigma=1\%$, $k=1\%$, $T_0=5$y, $T_1=6$y, $T_2=7$y. 
}     
\centering                                       
\begin{tabular}{cccc}                     
\hline\hline                                     
Bermudan Analytic & ${\rm Swaption}_{[T_1,T_2]}(0)$  &  ${\rm SprOpt}_{(2-1)}(0)$: Analytic  &  Berm PDE \\ [0.5ex]      
\hline                                                                        
0.004291718 & 0.0024225402 & 0.001869177   &   
0.00428736 \\
\hline\hline                                				
\end{tabular} \label{table:nonlin}  			
\end{table}

\begin{table}[hb] 
\caption{
The analytic result given in Eq.~(\ref{two_exercise_Berm}) is presented. 
}     
\centering                                       
\begin{tabular}{cccc}                    
\hline\hline    
  &  Lower Bound &    Upper Bound & ${\rm Analytic \; Bermudan }$  \\ [0.5ex]                             
\hline\hline  
K & ${\rm Swaption}[T_0,T_2]$  & ${\rm Swaption}[T_0,T_1]$  +  $[T_1,T_2]$ & ''${\rm Swaption}[T_0,T_1]$  +  $[T_1,T_2]$ +${\rm Corr}$''  \\ [0.5ex]      \hline                                                                     
$5\%$  & 0.00392194 &   0.00440466  & 0.004291718  \\  
$4\%$  & 0.00814275 &   0.00876597  & 0.008595882 \\ 
$3\%$  & 0.01498590 &   0.0156586 & 0.015446357 \\ 
$2\%$  & 0.02473200 &     0.0253359 & 0.025117529\\ 
$1\%$  & 0.03712100 &  0.0375695 & 0.037385166  \\  
\hline\hline                                				
\end{tabular} \label{table:nonlin}  			
\end{table}

\section{Three Exercise Dates Bermudan}

The three-exercise Bermudan price decomposes into three positive terms, and the structure of the successive additions of spread options as obvious further below, 
{\small
\bea
V( 0)  
%\hspace{-3mm}
&=& 
%\hspace{-3mm}
{\rm Swaption}_1(0) + {\rm SpreadOpt}_{(2-1)}(0) 
\\[3mm]\nonumber
&&\hspace{-8mm}
%\hspace{-3mm}
+
B(0,T_{-1})E^{T_{-1}} 
\left[
\Bigl(
\left. 
{\rm Swap}_3(T_{-1}) -  
{\rm Swaption}_1(T_{-1}) - {\rm SpreadOpt}_{(2-1)}(T_{-1}) \Bigr)^+
\right|{\cal F}_0
\right]
\;.
\label{ThreePointBerm}
\eea 
}
Here we have added another exercise date at $T_{-1}$ between time 0 and $T_0$.

\subsection{ Calculation of the larger Spread Option}

We analyze in this section each component of the last term of the three-exercise-date Bermudan swaption price.
 
\subsubsection{  Payer Swaption at $T_{-1}: $\; {\bf ${\rm Swaption}_{[T_1,T_2]}(T_{-1})$  } }

This swaption is equivalent to the one analyzed in Sect.~1, 
with the role of $T_0$ played by $T_{-1}$.

Rather than discretizing the evolution into small increments - using variables $y$ over $[T_{-1}, T_0]$ and $x$ over 
$[T_0, T_1]$ - we instead introduce a single Gaussian $Y$ that evolves the bond $ B(t; T_1, T_2)$ over the entire interval $[T_{-1}, T_1]$.

Define the boundary $Y_2^* (z) = D_{2-} + B_2 z$ and $Y_2^* (z) +\alpha_{2}(T_{-1}, T_1)  = D_{2+} + B_2 z$,
with $D_{2\pm}$, and $B_2 < 0$, as in the previous equation.
${\rm Swap}_3(T_{-1}) - {\rm Swaption}_{[T_1,T_2]}(T_{-1})$ yields the forward starting receiver swaption, 
as one of the components,
\bea
\label{Bondlets3_2}
&&
\hspace{-7mm}
{\rm \tilde{S}waption}\Bigl(T_{-1}; T_1,T_2,Y_2^*(z) \Bigr)   
%\hspace{-3mm}
\\[3mm]\nonumber
&&
=  
%\hspace{-3mm}
B(T_{-1},T_1) \left[(1+ \tau_2 K) B(T_{-1};T_1,T_2) 
N\Bigl( D_{2+} + B_2 z \Bigr) 
%\right.
%\nonumber\\[3mm]
%&&\left.
-
 N\Bigl( D_{2-} + B_2 z\Bigr) 
\right]
\quad.
%\label{Bondlets3_2}
\eea
$Y_2^*(z)$ denotes the argument of the receiver's cumulative distribution functions. Integration over $z$ yields terms $A_{2\pm} = D_{2\pm} - \alpha_{\pm} B_2$, with $\alpha_+ =  \alpha_{2}(T_{-1})$ and $\alpha_- =  \alpha_{1}(T_{-1})$.

\subsubsection{ Spread Option at $T_{-1}: $\; {\bf ${\rm SpreadOpt}_{(2-1)}(T_{-1})$  } }

An additional interval $z\in [0,T_{-1}]$ enters the forward bond evolution.
 For the spread ${\rm Swap}_{[T_0,T_1]}(T_0) -{\rm \tilde{S}waption}_{[T_1,T_2]}(T_0)$, the second term   vanishes at $x^*(y)$, while the spread admits a second zero $y^*(z)$, yielding two consecutive zeros.
Higher-order Bermudans introduce further zeros (e.g., $z^*(v)$) recursively.

The variable $x$ is Gaussian under the $T_1$-forward measure on $[T_0,T_1]$. Discounting uses the $T_0$-forward measure for $y$ and the $T_{-1}$-forward measure for $z$, under which both are Gaussian. The boundary 
  $x^*$ depends on $B(T_0;T_1,T_2)$, which depends on $y\in [T_{-1},T_0 ]$ and $z\in [0,T_{-1}]$. 

 As a consequence, $x^*$ depends on two random variables, $y$ and $z$, and this dependence is in fact linear.\footnote{ 
$x^*(y)$ depends on $B(T_{-1};T_1,T_2)$ which depends on $z$, i.e. 
$x^*(y)\bigr| + \alpha_2(T_0,T_1) =  d_{1\pm} + c_1 z  + b_1 y$; denote $d_{1\pm}(z) =  d_{1\pm} + c_1 z$, with
\bea
 d_{1\pm}
 &=& 
\frac{1}{\alpha_{2}(T_0, T_1)} 
\left[ 
 \ln \left( (1+\tau_2 K)\frac{B(0,T_2)   }{B(0,T_1)} \right)
 \pm \frac{1}{2} \alpha^2_{2}(T_0, T_1)
 \right.
\\[3mm]\nonumber
&&
\left.
-
 \frac{1}{2} \Bigl(\alpha^2_{2}(T_{-1},T_0) - \alpha^2_{1}(T_{-1},T_0)  \Bigr)
 -
 \frac{1}{2} \Bigl(\alpha^2_{2}(T_{-1}) - \alpha^2_{1}(T_{-1})  \Bigr)
\right]
\quad,
\label{d2_pm}
\eea
and
\be
c_1= -  \bigl(\alpha_{2}(T_{-1}) - \alpha_{1}(T_{-1})  \bigr) / \alpha_{2}(T_0, T_1)
\quad, \quad 
b_1= -  \bigl(\alpha_{2}(T_{-1},T_0) - \alpha_{1}(T_{-1},T_0)  \bigr) / \alpha_{2}(T_0, T_1)
\quad,
\ee 
both are scaled by $\alpha_{2}(T_0, T_1)$, with numerators given by differences of $\alpha$'s over the respective intervals. In Eq.~(\ref{SpreadOpt21}), $y$ and $z$ appear in the combined form $y+\frac{c_1}{b_1} z$, and $y+\frac{\alpha_1(T_{-1})-\alpha_0(T_{-1})}{\alpha_1(T_{-1},T_0)} z$. In the flat mean-reversion and flat volatility setting, both multpiplicative factors are equal, determining a strictly linear boundary $Y^*_1(z)$. 
}
This zero enters into the cumulative Gassian of the receiver!

Using this, we value the spread option, whose solution defines
$Y^*_1(z)$,
\footnote{
More generally, the damping factors attached to the exponential terms preserve near-linearity of the boundary.}
\bea
 \label{SpreadOpt21}
&&
\hspace{-6mm}{\rm SpreadOpt}_{(2-1)}(T_{-1})  
%\hspace{-3mm} 
=
%\hspace{-3mm} 
B(T_{-1},T_0)
E^{T_0} \left[
 \left\{ 1-
\Bigl( (1+ \tau_1 K) 
\right.\right.
\\[3mm]
&&
\left.
\left.
\hspace{-6mm}- N(d_{1-}(z)  + b_1 y) \Bigr)  \frac{ B(T_{-1},T_1) }{ B(T_{-1},T_0) } 
 \exp\Bigl\{ - \alpha_1(T_{-1},T_0) y - \frac{1}{2} \alpha^2_{1}(T_{-1},T_0)    \Bigr\}
\right.\right.
\nonumber\\[3mm]\nonumber
&&\left.\left.
\left.
\hspace{-6mm}-(1+ \tau_2 K) 
\frac{ B(T_{-1},T_2) }{ B(T_{-1},T_0) } 
\exp\left\{  - \alpha_2(T_{-1},T_0) y - \frac{1}{2} \alpha^2_{2}(T_{-1},T_0)     \right\} 
N\Bigl(d_{1+}(z) + b_1  y\Bigr) 
\right\} ^+ \right|{\cal F}_{-1}\right]
\;.
% \label{SpreadOpt21}
 \eea

As variance accumulates in $d_{1\pm}$, their values tend to shift further into the negative region. Consequently, the cumulative distribution functions increasingly suppress the exponential factors in terms such as $N(d_{1\pm}(z))\exp\{-\alpha_{\pm} z\}$. 
This reinforces the near-analytic valuation of the 
the next zero function 
as one advances to higher-point Bermudans.

After integrating over $y$ in the spread equation, we switch from $d_{1\pm}(z)$ to $a_{1\pm}(z)$.
From Eq.~(\ref{TT_integrals_0}) we have 
$a_{1\pm} = d_{1\pm} - \alpha_\pm b_1, \alpha_+ = \alpha_2(T_{-1},T_0), \alpha_- = \alpha_1(T_{-1},T_0)$.

The first term in (\ref{SpreadOpt21}) yields ${\textit Swaption}_{[T_0,T_1]}(T_{-1})$, while the second term constitutes the integral over $y$ of the forward starting receiver swaption in $[T_1,T_2]$, 
\bea
{\rm SpreadOpt}_{(2-1)}(T_{-1};z) &=& 
{\rm Swaption}\Big(T_{-1};T_0,T_1\Bigr) \Bigl( Y_1^*(z) \Bigr)  
\\[3mm]\nonumber
&-& B(T_{-1},T_0)\int_{Y_1^*(z)}^{\infty} {\rm \tilde{S}waption}\Bigl(T_{0};T_1,T_2, x^*(y,z)\Bigr) \; \phi(y) \rd y
\quad.
\label{Spread_2minus1_02}
\eea

\subsubsection{ Calculation of: \\
%{\small
$
B(0,T_{-1})E^{T_{-1}} 
\left[
\Bigl(
\left. 
{\rm Swap}_3(T_{-1}) -  
{\rm Swaption}_1(T_{-1}) - {\rm SpreadOpt}_{(2-1)}(T_{-1}) \Bigr)^+
\right|{\cal F}_0
\right]
$ 
%}
}
\vspace{3mm}

Here we use the put-call parity to obtain 
\bea
%\nonumber
&&
{\rm Swap_3}(T_{-1}) - {\rm Swaption}_{[T_1,T_2]}(T_{-1}) - {\rm Swaption}_{[T_0,T_1]}(T_{-1}) 
\\[3mm] \nonumber
&&
\hspace{20mm}
=  {\rm Swap}_{[T_{-1},T_0]} - {\rm \tilde{S}waption}_{[T_0,T_1]}(T_{-1}) - {\rm \tilde{S}waption}_{[T_1,T_2]}(T_{-1})
\;.
\eea
Taking the expectation we get,
\bea
\label{Spread_3_21_1}
&&\hspace{-10mm}
{\rm SpreadOption}_{[3-1-(2-1)]}(0) 
  =  {\rm Swaption}_{[T_{-1},T_0]} (z^*)  
\\[3mm]\nonumber
&&
-B(0,T_{-1}) \sum_{j=1,2}\int_{z^*}^\infty{\rm \tilde{S}waption}\Bigl(T_{-1};T_{j-1},T_j,Y^*_j(z)\Bigr) \phi(z)\rd z 
\\[3mm]\nonumber
&& 
+ B(0,T_{-1}) \int_{z^*}^\infty \rd z \phi(z) B(T_{-1},T_0) \int_{Y_1^*(z)}^\infty {\rm \tilde{S}waption}\Bigl(T_{0};T_{1},T_2, x^*(y,z) \Bigr) \phi(y) \rd y 
\quad.
\eea
At each exercise date, the corresponding short swaption is added, and all subsequent forward-starting short receiver swaptions from that exercise date onward are subtracted.

\subsection{The three-exercise Bermudan price}

\begin{figure}[H] 
\center
{
\includegraphics[width =0.34\textwidth]{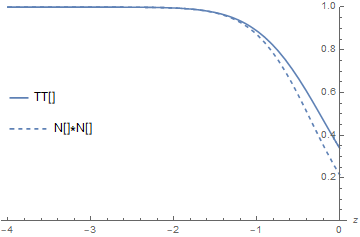} 
\hspace{15mm}
\includegraphics[width =0.35\textwidth]{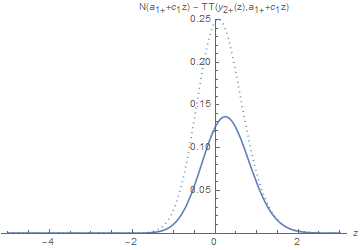} 
}
\caption{ 
On the left, we show the large negative-$z$ approximation of the $TT$-function, last line in Eq.~(\ref{3PointBerm}).
On the right, we display the asymptotic behavior, for large $|z|$, of each square bracket. 
The dotted line approximates them well for $|z|\gg 0$.
Near the ATM point, the positive and negative contributions are non-zero and slightly asymmetric, but their net effect is negligible.
 }
\label{N_TT_Functions}
\end{figure}

A closed-form expression for the price of the three-exercise Bermudan option is given by:
{\small
\bea
\label{3PointBerm}
V^{(3)}(0)
\hspace{-2mm}
&=& 
\hspace{-2mm}
V^{(2)}(0)
+B(0,T_{-1}) \left[ N(-z^*) - \frac{B(0;T_{-1},T_0)}{\tilde{K}_0} N(-z^* -\alpha_0(T_{-1}) ) \right] 
\\[3mm]\nonumber
&-&
\hspace{-3mm}
\sum_{j=1,2}B(0,T_{j-1})\left\{
\frac{ B(0;T_{j-1},T_{j}) }{ \tilde{K}_{j} }
\left[ N\left(\frac{ {A}^*_{j+}}{\sqrt{1+B^2_{j}}} \right) 
-
 TT\bigl(z^* + \alpha_j(T_{-1}), {A}^*_{j+} \bigr)
\right]
\right.
\\[3mm]
&&
\left.
-
\left[
N\left(\frac{A^*_{j-}}{\sqrt{1+B_j^2}} \right) 
-
TT\bigl(z^* + \alpha_{j-1} ,  {A}^*_{j-} \bigr)
\right]
\right\}
\nonumber\\[3mm]\nonumber
&+&
\hspace{-2mm}
B(0,T_{-1}) 
\int_{z^*}^\infty
\rd z \phi(z)
\left\{
\frac{B(T_{-1},T_2)}{\tilde{K}_2}\left[ N\left(\frac{\tilde{x}^*_{+}(z)}{\sqrt{1+b_1^2}}\right) - TT \Bigr(Y_{1+}^*, \tilde{x}^*_{1+}(z)\Bigl)\right] 
\right.
\nonumber\\[3mm]\nonumber
&&
\left.
-
B(T_{-1},T_1)\left[N\left(\frac{\tilde{x}^*_{-}(z)}{\sqrt{1+b_1^2}}\right) - TT \Bigl(Y_{1-}^*, \tilde{x}^*_{-}(z)\Bigr)\right]
\right\}
\quad.
\eea
}
$V^{(2)}(0)$ is calculated in (\ref{two_exercise_Berm}). 
The solution for $z^*$ is obtained from Eq.~(\ref{z*_Tminus1_1}).\footnote{
Equation for $z^*$, solution of ${\rm Swap_3}(T_{-1}) - {\rm Swaption}_{[T_1,T_2]}(T_{-1}) - {\rm SpreadOpt}_{(2-1)}(T_{-1}) = 0$, 
\bea
\label{z*_Tminus1_1}
&&
\hspace{-20mm}
{\rm Swap_3} - {\rm Swaption}_{[T_1,T_2]} - {\rm SpreadOpt}_{(2-1)} = 
1 - (1 + \tau_0 K)  B(T_{-1},T_0) 
\\[3mm]
&-&
\sum_{j=1,2} 
(1+\tau_j K)B(T_{-1},T_j) N\Bigl( D_{j+} + B_j z \Bigr) -  B(T_{-1},T_{j-1}) N\Bigl( D_{j-} + B_j z \Bigr)
\nonumber\\[1mm]
&+& 
%\hspace{-5mm}
(1+\tau_2 K) B(T_{-1},T_2) 
\left[ N\left(\frac{a_{1+}(z)}{\sqrt{1+b_1^2}}\right) - TT \Bigl( Y_1^*(z) +\alpha_2(T_{-1},T_0), a_{1+}(z)\Bigr)
\right] 
\nonumber\\[3mm]\nonumber
&-&
B(T_{-1},T_1) 
\left[N\left(\frac{a_{1-}(z)}{\sqrt{1+b_1^2}}\right) - TT \Bigl(Y_1^*(z) +\alpha_1(T_{-1},T_0),a_{1-}(z)\Bigr) 
\right]
\quad.
\eea
}
In this expression, the two terms in the first line are supplemented by six exponentials with damping terms of the form $N(d_{i\pm} + b_i z)$. The damping in the last line is significantly stronger (Fig.~\ref{N_TT_Functions}). 

Backward induction introduces optionality-switch corrections to the Bermudan, with each added swaption offset by a much smaller forward-starting receiver swaption integrals.
 
The last term is still expressed as a product of exponential functions and cumulative distribution functions of linear functions of 
$z$. 
 Owing to the linear dependence on $z$, the two arguments of the $TT$ functions possess distinct zeros, but it can be proven that the term is well-behaved.

The boundary $Y^*_1(z)$ is exactly linear. The constant part can be found numerically with Newton-Raphson, but can be approximated analytically.\footnote{
For the $y$-integration, from $b_1$, we have $\alpha_\pm = \alpha_{2/1}(T_{-1},T_0)$ and $Y_1^*+ \alpha_{\pm} = Y^*_{1}(\pm)$. For the $z$-integration and $B_1$ we have $\alpha_+ = \alpha_1(T_{-1})$ and $\alpha_- = \alpha_0(T_{-1})$, 
yielding
\bea
Y_1^* = D_{1-} + B_1 z \; , \; Y_1^* +\alpha_1(T_{-1},T_0) = D_{1+} + B_1 z
\;, \; 
Y_1^* +\alpha_2(T_{-1},T_0) = D^+_{1-} + B_1 z \;, \;
D^+_{1-}  = D_{1-} + \alpha_2(T_{-1},T_0)
.
\nonumber
\eea
$Y^*_1(z)$ admits an analytical approximation, whereas $ Y_2^* (z)$ is obtained in closed form; below $j=1,2$ approximation applied only to $Y^*_1(z)$,
\bea
 Y_j^* (z)
 &=& 
\frac{1}{\alpha_{j}(T_{-1}, T_{j-1})} 
\left[ 
 \ln \left(  \frac{B(0;T_{j-1},T_{j})}{\tilde{K}_j} \right)
-
 \frac{1}{2} 
  \alpha^2_{j}(T_{-1}, T_{j-1})
\right]
%\nonumber\\[3mm]
\\[3mm]\nonumber
 &-&
 \frac{1}{2}
\left( \frac{ \alpha^2_{j}(T_{-1}) - \alpha^2_{j-1}(T_{-1}) }{
\alpha_{j}(T_{-1}, T_{j-1})
}
\right)
- 
\left( \frac{ \alpha_{j}(T_{-1}) - \alpha_{j-1}(T_{-1})  }{\alpha_{j}(T_{-1}, T_{j-1}) }\right) z 
\quad,
\label{yj_z}
\eea
and at the integrated $z$-terms, $A_{j\pm} = D_{j\pm} - \alpha_{j\pm} B_j$, where $D_{j\pm}$ and 
$B_{j\pm}$ can be red off the above.
}

Both gradients $B_1$ and $c_1$ of the last line $TT$ functions are negative. Eq.~(\ref{Large_h_Large_a}) implies that, for large negative $z$, the surviving term of the $TT$ function is $N(D^+_{1-} + B_1 z ) N\left(\frac{a^*_{1+} + c_1 z}{\sqrt{1+b_1^2}}\right)$.
The difference with the cumulative distribution function, first term in the square bracket, yields
$
 N(-D^+_{1-} - B_1 z ) N\left(\frac{a^*_{1+} + c_1 z}{\sqrt{1+b_1^2}}\right)
 $, for $z\ll 0$.
For large positive $z$, the $TT$ function decays rapidly, leaving only the cumulative term, which also tends to zero. The expression captures well the asymptotic behavior for large positive and negative $z$, where the function vanishes in both directions. This remains true despite the exponential prefactors.

\begin{figure}[H] 
\center
{
\includegraphics[width =0.45\textwidth]{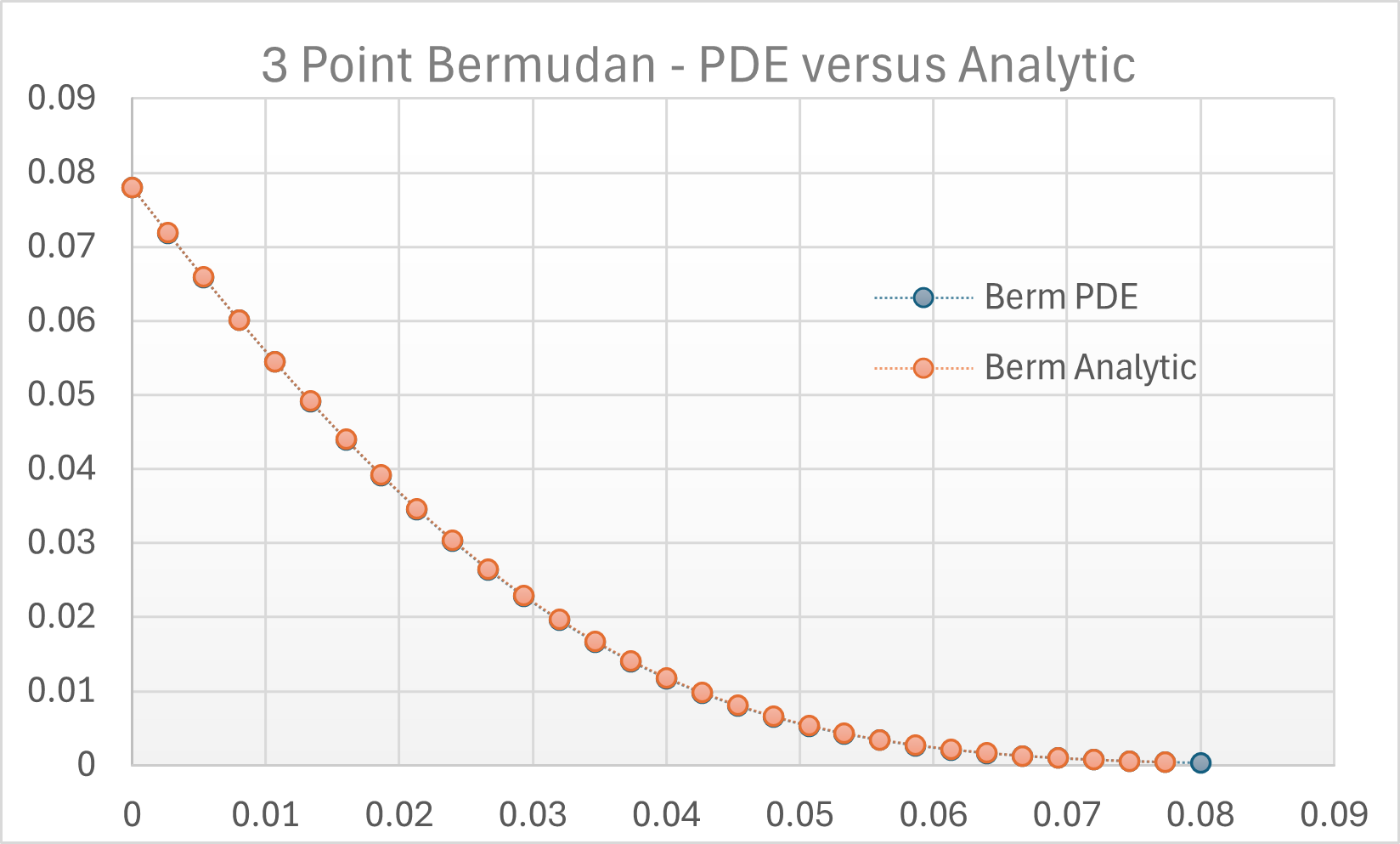}
\includegraphics[width =0.45\textwidth]{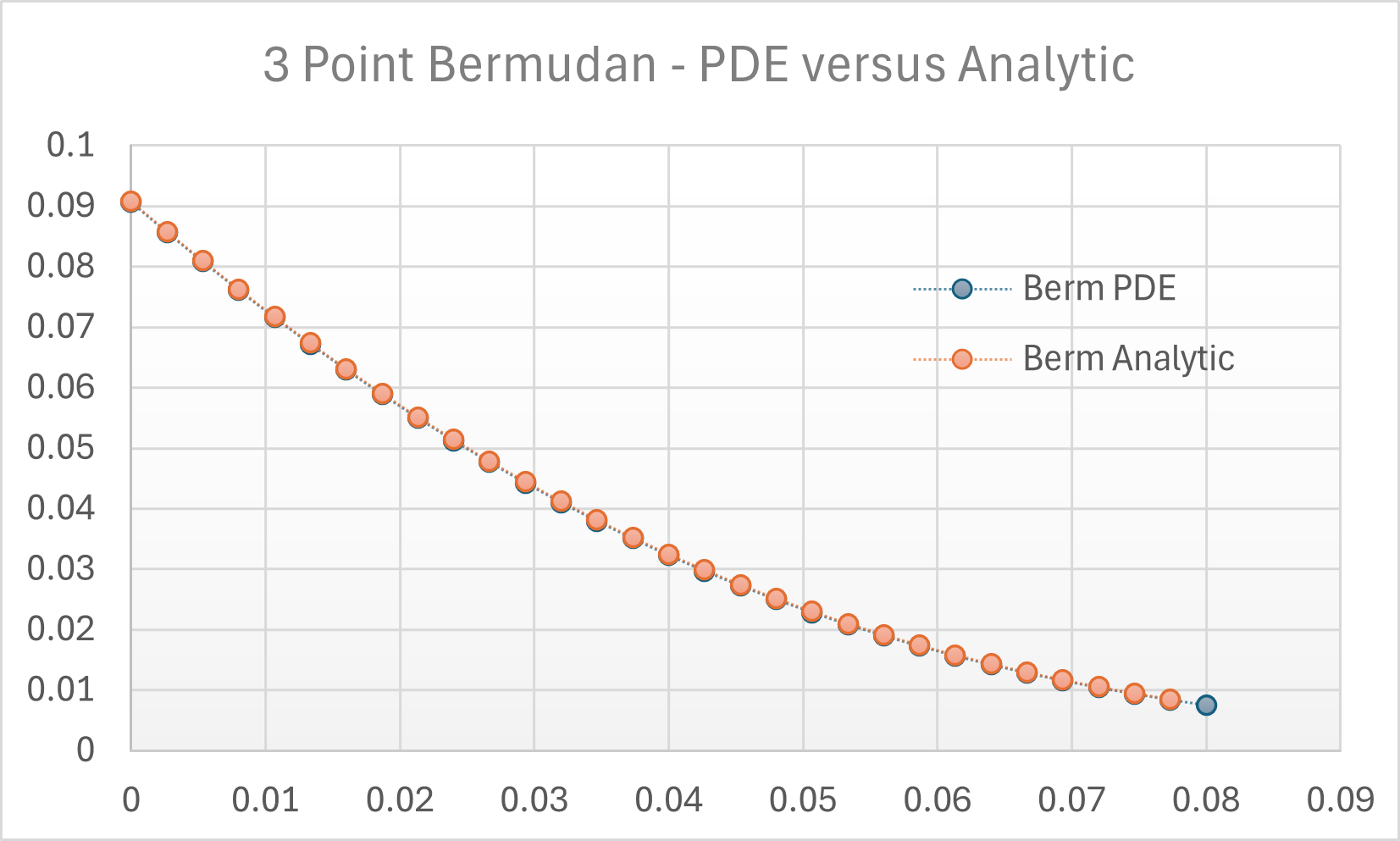}
}
\caption{ PV three-point Bermudan option on an annual Swap starting at $T_{-1}=4$y, ending at $T_2=7$y, with three exercise dates for a range of strikes, is shown.  The volatility is taken to be $\sigma=1\%$ on the left and $\sigma=2\%$.
}
\label{Three_Point_Berm}
\end{figure}

\begin{table}[hh]
\caption{Analytic Bermudan, (\ref{3PointBerm})\&(\ref{yj_z}). The parameters are $r=3\%$, $\sigma=1\%$, $k=1\%$, $T_{-1}=4$y, $T_0=5$y, $T_1=6$y, $T_2=7$y. }    
\centering                                       
\begin{tabular}{cccccc}                    
\hline\hline                                     
K & Bermudan & {\small ${\rm Swaption}_{[T_1,T_2]}$} & $[T_0,T_1]$  &$-\int \tilde{S}(T_{0};{T_1,T_2})$ & $[T_{-1},T_0]$ \\ [0.5ex]      \hline                                                            
$5\%$ & 0.00564056 & 0.00242254 & 0.00198212 & -0.000112938 & 0.00144684 \\  
$4\%$ & 0.0118261 & 0.00464554 & 0.00412043 &  -0.000170088  & 0.00341415 \\ 
$3\%$ & 0.0220608 & 0.00806763 & 0.00759102 &  -0.000212243  & 0.00689059  \\ 
$2\%$ & 0.0368763 & 0.0127984 & 0.0125375 &  -0.000218371  & 0.0120889 \\ 
$1\%$ & 0.0558895 & 0.0187416 & 0.0188278 &  -0.000184334  & 0.0188165  \\  
\hline\hline                                				
\end{tabular} 
\label{table:3_Point_Berms}  			
\end{table}

\begin{table}[hh]
\caption{Analytic Bermudan, (\ref{3PointBerm})\&(\ref{yj_z}). The parameters are $r=3\%$, $\sigma=1\%$, $k=1\%$, $T_{-1}=4$y, $T_0=5$y, $T_1=6$y, $T_2=7$y. (Continued from Table \ref{table:3_Point_Berms}).  }    
\centering                                       
\begin{tabular}{ccc}                    
\hline\hline                                     
K &  $-\int \tilde{S}(T_{-1};{T_0,T_1})$  & -$\int \tilde{S}(T_{-1};{T_1,T_2})$ \\ [0.5ex]      \hline                                                            
$5\%$ &  -0.0000641896 & -0.0000622925\\  
$4\%$ &  -0.000111855 & -0.00010855\\ 
$3\%$ &  -0.000154562 & -0.000149994\\ 
$2\%$ &  -0.000168369 & -0.000163393\\ 
$1\%$ &  -0.000143746  & -0.000139498\\  
\hline\hline                                				
\end{tabular} 
\label{table:3_Point_Berms_02}  			
\end{table}

\begin{table}[hb] 
\caption{
Analytical vs. numerical comparison. Analytical results from Eq.~(\ref{3PointBerm}) and PDE solutions (first column) are reported; the final column is computed from Eq.~(\ref{SpreadOpt21}) for $Y_1^*(z)$, and the penultimate column uses the approximation in Eq.~(\ref{yj_z}).
}     
\centering                                     
\begin{tabular}{cccccc}                   
\hline\hline    
  & %Bermudan 
  PDE & Lower Bound &    Upper Bound & ${\rm Analytic \; Berm }$ & ${\rm Analytic \; Berm }$ \\ [0.5ex]                                   %
\hline\hline  
K & Bermudan & {\small ${\rm Swpt}_{[T_{-1},T_2]}$}  &  
{\small $\sum_{j=0}^2{\rm S}_{[T_{j-1},T_j] }$} %+[T_0,T_1]+[T_1,T_2]}$}
 & {\small ``Eq.~(\ref{3PointBerm})\&(\ref{yj_z})''} & {\small ``Eq.~(\ref{3PointBerm})\&(\ref{SpreadOpt21})} 
\\ [0.5ex]      \hline                                                                      
$5\%$ & 0.00557887 & 0.00441743 &   0.0058515  & 0.00564056 & 0.00564057 \\  
$4\%$ & 0.0117321 & 0.0102559 &   0.0121801  & 0.0118261 & 0.0118262 \\ 
$3\%$ & 0.0219506 & 0.0204681 &   0.0225492 & 0.0220608 & 0.0220765 \\ 
$2\%$ & 0.03678 & 0.0355979 &     0.0374247 & 0.0368763 & 0.0368767\\ 
$1\%$ & 0.0558436 & 0.0550954 &  0.056386 & 0.0558895  & 0.0558895 \\  
\hline\hline                                				
\end{tabular} \label{table:BermPrices}  			
\end{table}

\section{General Case }

We extend our notation to denote all Gaussian variables as $y_i$, $i = 0, \ldots, n-1$. 
There are $n+1$ time intervals and $n$ corresponding Gaussian variables. No Gaussian is associated with the last interval. If the rapidly decaying double and higher-order integrals are neglected, the Bermudan option can be reproduced by the following:
\bea
\label{Low_Exercise_0}
V^n( 0) &=& V^{n-1}( 0) +
{\rm Swaption}_{[T_0,T_{1}]}(y_0^*) 
\\[3mm]\nonumber
&-& B(0,T_{0}) \sum_{j=1}^{n-1} \int_{y^*_0}^\infty {\rm \tilde{S}waption}\Bigl( T_0;T_j,T_{j+1}, Y^*_j(y_0) \Bigr) \phi(y_0) \rd y_0
\quad,
%\label{Low_Exercise_0}
\eea
with $Y_j^* (y_0) = d_{\pm} (T_0;T_j,T_{j+1}) + b(T_0;T_j,T_{j+1}) y_0$, 
generalisation of Eq.~(\ref{yj_z})
\bea
 Y_j^* (y_0)
 &=& 
\frac{1}{\alpha_{j+1}(T_0, T_j)} 
\left[ 
 \ln \left(  \frac{B(0;T_{j},T_{j+1})}{\tilde{K}_{j+1}} \right)
-
 \frac{1}{2} 
  \alpha^2_{j+1}(T_{0}, T_j)
\right]
\nonumber \\[3mm]
 &-&
 \frac{1}{2}
\left( \frac{ \alpha^2_{j+1}(T_{0}) - \alpha^2_{j}(T_{0}) }{
\alpha_{j+1}(T_{0}, T_j)
}
\right)
- 
\left( \frac{ \alpha_{j+1}(T_{0}) - \alpha_{j}(T_{0})  }{\alpha_{j+1}(T_{0}, T_j) }\right) y_0 
\;,
\label{y1_z}
\eea
with $y^*_0$ obtained by solving an equation analogous to (\ref{z*_Tminus1_1}), typically solved by Newton-Raphson, with an analytical initial guess obtained by neglecting exponential terms with damping factors.

The Bermudan option can be reconstructed by adding, at each exercise date, the initial short swaption and subtracting all forward-starting receiver swaptions that commence at that date. 
The ensuing double and higher-order integrals decay rapidly and, when only a limited number of exercise dates is involved, may be neglected without materially affecting the valuation.

Although the strikes of the swaptions increase due to the presence of subsequent receiver swaptions, they may be approximated by the Bermudan strike for practical purposes. With that approximation we have
{\small
\bea
\label{3PointBerm_2}
V^n(0)
\hspace{-2mm}
&=& V^{n-1}(0)+ 
B(0,T_{0}) \left[ N(-y_0^*) - \frac{B(0;T_{0},T_{1})}{\tilde{K}_{1}} N\Bigl(-y_0^* -\alpha_{1}(T_{0}) \Bigr) \right] 
\\[3mm]\nonumber
&-& 
\sum_{j=1}^{n-1} B(0,T_j)
\left\{
\frac{ B(0;T_{j},T_{j+1}) }{ \tilde{K}_{j+1}}
\left[ N\left(\frac{A_{j+}}{\sqrt{1+B^2_{j}}} \right) 
-
 TT\Bigl(y_0^* + \alpha_{j+1}(T_{0}), A_{j+} \Bigr)
\right]
\right.
\\[3mm]\nonumber
&&
\left.
-
\left[
N\left(\frac{A_{j-}}{\sqrt{1+B^2_{j}}} \right) 
-
TT\bigl(y_0^* + \alpha_j(T_{0}) ,  A_{j-} \bigr)
\right]
\right\}
\quad.
\eea
}
Also $A_{j\pm}=A_{\pm} (T_0;T_j,T_{j+1})= D_{\pm} (T_0;T_j,T_{j+1})\pm \alpha_{j+1/j}(T_0,T_j) b(T_0;T_j,T_{j+1})$, see (\ref{z*_Tminus1_1}). 

One can extend the calculations to more exercise points and to higher order integrals, %{\small
\bea
\label{Low_Exercise_2}
&&V^n( 0) = V^{n-1}( 0) +
{\rm Swaption}_{[T_0,T_{1}]}(y_0^*) 
- B(0,T_{0}) \sum_{j=1}^{n-1} \int_{y^*_0}^\infty 
\rd y_0 \phi(y_0) \times
\\[3mm]\nonumber
&& 
\times \left[
{\rm \tilde{S}waption}\Bigl( T_0;T_j,T_{j+1}, Y^*_{j}(y_0) \Bigr) 
\right.
\\[3mm]\nonumber
&&
\left.
-\sum_{k=1}^{n-j-1} B(T_0,T_j) \int_{Y^*_{j}(y_0)}^\infty
{\rm \tilde{S}waption}\Bigl( T_j;T_{j+k},T_{j+1+ k}), y^*_{[T_j, T_{j+k}]}(Y_{j}) \Bigr) \phi(Y_{j}) \rd Y_{j}
\right]
\;.
%\label{Low_Exercise_2}
\eea

The pattern of the terms in the pricing of the Bermudan follows same idea. For every added swaption ${\rm Swaption}_{[T_0,T_{1}]}$, all forward starting swaptions ${\rm \tilde{S}waption}\Bigl( T_0;T_j,T_{j+1}\Bigr)$  over all remaining time intervals $[T_j,T_{j+1}]$ are subtracted, and for each subtracted forward swaptions, integrals of forward starting swaptions on the remaining intervals are added, and this process continues.

The cumulative Gaussians in the ${\rm \tilde{S}waption}$'s contain
\be
\label{gradients}
N\left(y_i\right) = N\left(d_{i,-} +  \sum_{j=0}^{i-1}\epsilon^i_j y_j \right)   
\quad, \quad 
\epsilon^i_j 
= -
 \frac{\Delta \alpha_{i+1}\bigl({T_{j-1},T_{j}} \bigr)  
}{\alpha_{i+1}(T_{i-1},T_{i})}
%\quad,\quad
%i=0,\ldots, n-1\quad, \quad j=0, \ldots, i-1
\quad,
\ee
where $i=0,\ldots, n-1$, $j=0, \ldots, i-1$, and the first term constants $ d_{i+1,\pm}^*$ are given through analytical approximation by (with $T_{-1}\equiv 0$) 
\bea
d_{i,\pm} 
&=&
\frac{1}{\alpha_{i+1}(T_{i-1}, T_{i})} 
\left[ 
 \ln \left( \frac{ B(0;T_{i},T_{i+1})   }{ \tilde{K}_{i+1 } } \right)
\pm  \frac{1}{2}\alpha^2_{i+1}(T_{i-1},T_{i})
\right]
\nonumber\\[3mm]
&-&
\frac{1}{2} \sum_{j=1}^{i}
\left( \frac{ \alpha^2_{i+1}\bigl({T_{i-1-j},T_{i-j}} \bigr) 
-
\alpha^2_{i}\bigl({T_{i-1-j},T_{i-j}} \bigr) 
}{\alpha_{i+1}(T_{i-1},T_{i})}
\right)
\quad.
\eea
 For every Gaussian $i$, there are $i$ gradients $\epsilon_j^i$. 
In the above, $\Delta \alpha_{i+1} = \alpha_{i+1}\bigl(T_{j-1},T_{j}  \bigr) 
- \alpha_{i}\bigl(T_{j-1},T_{j} \bigr)$.

In (\ref{gradients}), the approximation for calculating $d_{i\pm}$ relies on the additional damping in the exponential terms, which allows them to be neglected, alternatively a Newton-Raphson should be used. The remaining terms correspond to standard swaptions and forward-starting swaptions with the same strike as the Bermudan, or higher in the case of numerical Newto-Raphson caclulation.


\begin{thebibliography}{9}

\bibitem{AndersenPiterbarg2010}
 {Andersen, Leif B. G. and Piterbarg, Vladimir V.},
  {Interest Rate Modeling, Volume I: Foundations and Vanilla Models},
  %{Term Structure Models},
  %{Products and Risk Management},
 {Atlantic Financial Press},
 {2010},
 {London}

\bibitem{AndersenPiterbarg2010_II}
 {Andersen, Leif B. G. and Piterbarg, Vladimir V.},
  {Interest Rate Modeling, Volume II: } %Foundations and Vanilla Models},
  {Term Structure Models},
  %{Products and Risk Management},
 {Atlantic Financial Press},
 {2010},
 {London}

\bibitem{AndersenPiterbarg2010_III}
 {Andersen, Leif B. G. and Piterbarg, Vladimir V.},
  {Interest Rate Modeling, Volume III: }
  %Foundations and Vanilla Models},
  % {Term Structure Models},
  {Products and Risk Management},
 {Atlantic Financial Press},
 {2010},
 {London}



\bibitem{Henrard} {M. Henrard} , {Semi
Explicit Approach to Canary Swaptions in HJM One-Factor Model
Applied Mathematical Finance} , {Volume 13 , Issue 1 , p. 1 - 18 Posted: 2006-03}

\bibitem{Feldman} {K. Feldman}, {Berms without Calibration}, {The Journal of Risk, 2025[10.21314/JOR.2025.015]}


\bibitem{OwenT}
  {Owen, Donald B.},
  \textit{Tables for Computing Bivariate Normal Probabilities},
  {Annals of Mathematical Statistics},
   Volume  {27},
  Nr.   {4},
  pgs {1075--1090},
  {1956}.
%  , doi {10.1214/aoms/1177728074}
  


\end{thebibliography}
\end{document}